\documentstyle[aps,graphics,multicol,psfig,epsfig,color]{revtex}  

\begin{document}

\title{Diffusion, peer pressure and tailed distributions}

\author{Fabio Cecconi$^{\ddag}$, Matteo Marsili$^{\ddag}$,
Jayanth R. Banavar$^{\dag}$ \& Amos Maritan$^{\ddag}$ }

\address{$^{\ddag}$ International School for Advanced Studies (S.I.S.S.A.),
Via Beirut 2-4, 34014 Trieste, INFM and the Abdus Salam International
Center for Theoretical Physics, Trieste, Italy \\
$^{\dag}$ Department of Physics,
104 Davey Laboratory, The Pennsylvania State University, University
Park, Pennsylvania 16802}

\date{\today}  

\maketitle 

\begin{abstract}
We present a general,
physically motivated non-linear and non-local advection equation in
which the diffusion of interacting random walkers competes with a
local drift arising from a kind of peer pressure. 
We show, using a mapping to an integrable
dynamical system, that on varying a parameter, the steady state
behaviour undergoes a transition from the standard diffusive behavior
to a localized stationary state characterized by a tailed
distribution. Finally, we show that recent empirical laws on economic
growth can be explained as a collective phenomenon due to peer
pressure interaction.
\end{abstract}


\begin{multicols}{2}
\narrowtext           

Fluctuations -- measured by deviations from the mean value of an
observable -- of large systems often satisfy a Gaussian distribution.
A classic example is a linear diffusion process
\cite{RW1,RW2,RW3,Chandra} which has numerous applications in many
branches of science \cite{SWM}.  However, there are many situations in
Nature in which the probability of occurence of a fluctuation of size
$|\Delta|$ is proportional to $\exp(-|\Delta|^p)$, with $p$ taking on
a value greater than or equal to $1$ (exponential tails), but smaller than
$2$ \cite{Leo}.  Examples include the temperature distribution of a
Rayleigh-Benard system \cite{RB,CH}, disordered systems such as foams
\cite{foams} and glasses or granular materials \cite{Granular}, with even 
fatter tails in financial data \cite{Mantegna,Finance}.  

We present here a physically motivated 
advection equation and its exact steady state solution.  The equation has 
a drift term, originating from a 
kind of a peer pressure,
of the same nature as that due to mechanisms of chemotactic signalling
by microorganisms \cite{Chemo,Murray} or by the onset of cooperation in social
groups \cite{Axel,SigNow98}, or by competition between economic
units \cite{Stanley}. We show that, on varying a parameter, there is a 
transition
from diffusive behavior to a localized stationary state characterized by 
an exponential distribution.

Consider a diffusional process in which random walkers move either to the
right or to the left randomly and with no bias.
In the long time limit, the distribution of walkers becomes flat and
infinitely spread out. The mechanism for collective self-organization in
our model is a kind of peer pressure.
The non-linearity arises from an interaction (of spatial range
$\xi$)  between the walkers,  which leads to a drift or a bias
term which opposes the diffusional spreading and promotes
aggregation.  The basic idea is one in which a
walker perceives the populations of other walkers over a range
$\xi$ both right and left of her own location and has a drift in
the more crowded direction.  In the limit of
small $\xi$, one obtains regular diffusive behavior, whereas a
new class of steady state behavior characterized by non-Gaussian
distributions is obtained when $\xi$ is sufficiently
large.

This idea may be encapsulated in a nonlinear 
equation
\begin{equation}
\frac{\partial P}{\partial t} =
- \frac{\partial}{\partial x} \bigg(
v(x,t) P \bigg)
+ D\frac{\partial^2 P}{\partial x^2}
\label{eq:Pdot}
\end{equation}
where $P(x,t)$ is the distribution function of the
locations of the random walkers at time $t$ with a drift
velocity $v(x,t)$.  The nonlinearity arises because $v$ is itself a
non-local function of $P$ and is given by
\begin{equation}
v(x,t) = \frac{\lambda}{2}\;
\frac{\Phi_{+}(x,t) - \Phi_{-}(x,t)}{\Phi_{+}(x,t) + \Phi_{-}(x,t)}
\label{eq:v}
\end{equation}
where
\begin{eqnarray}
 \Phi_{+}(x,t) =
 \int_x^{\infty} dy e^{(x-y)/\xi} P(y,t)\;, \label{eq:phip} \\
 \Phi_{-}(x,t) =
 \int_{-\infty}^x dy e^{(y-x)/\xi} P(y,t)\;, \label{eq:phim}
\end{eqnarray}
and $\lambda$ sets the scale for the drift velocity.  Physically,
$\Phi_{+} (x,t)$ and $\Phi_{-} (x,t)$ are measures of the population,
within a range $\xi$, to the right and left of location $x$ and the
drift velocity is then a normalized imbalance between these two
populations.
 
In colonies of social individuals, each
member can release chemical substances in the environment
(for example, pheromones) revealing the presence of food sources in
a given area.
The individuals in the colony are able to detect higher concentrations of the
chemical signal and are attracted toward the food and move accordingly.
The plausible assumption that the concentration of the
chemicals in a given region is proportional to the local density of
individuals translates into an effective interaction between members of the 
colony, and leads to the nonlocal drift term in Eq.~(\ref{eq:Pdot}).
The interaction causes the net migration of an individual in the direction
of the higher local density.
The length scale, $\xi$, is a measure of the range of the
biological sensory system of individuals.  Thus
Eq.~(\ref{eq:Pdot}) describes Brownian motion with a
bias that  mimicks attractive chemotactic signalling \cite{Chemo} and 
promotes aggregation leading to a drift controlled by the population 
difference in localized regions on the right and left of $x$.
Physically, the standard Gaussian solution is obtained when
there is no drift ($\lambda = 0$) or when the interaction length scale goes
to zero ($\xi \to 0$).

The existence of a transition in the steady state behavior is revealed
by a linear stability analysis of the homogeneous state.  One can look
for solutions of Eq.~(\ref{eq:Pdot}) in the form $P \sim \exp [ikx -
\gamma(k)t]$ representing spatially periodic perturbations.  A direct
substitution into Eq.~(\ref{eq:Pdot}) provides the exponential growth
rate

\begin{equation}
\gamma(k) = D(1-r) k^2 \;,
\end{equation}
where
\begin{equation}
r = \frac{\lambda \xi}{2D}\;.
\label{eq:r}
\end{equation}

Hence the homogeneous state is stable ($\gamma(k)>0$) when $r < 1$.
For $r>1$ on the contrary, small perturbations grow with time and the
homogeneous state becomes unstable. Thus $r$ plays the role of a
control parameter of a phase transition.

A mapping to an Hamiltonian dynamical systems allows us to derive the
stationary soltuion of Eq.~(\ref{eq:Pdot}). Indeed the steady state
distribution ($\dot{P_s} = 0$) satisfies the ordinary differential
equation 

\begin{equation} 
P_s' = \frac{\lambda}{2D} \frac{\Phi_{+} - \Phi_{-}}{\Phi_{+} +
\Phi_{-}} P_s
\label{eq:statio}
\end{equation}
where the prime denotes the derivative $d/dx$.
The steady state values of $\Phi_{+}$ and $\Phi_{-}$, in turn,
satisfy
\begin{eqnarray}
\Phi_{+}' &=&  \Phi_{+}/\xi - P_s \label{eq:plus} \\
\Phi_{-}' &=& -\Phi_{-}/\xi + P_s \label{eq:minus}
\end{eqnarray}
obtained by taking derivatives of Eqs.~(\ref{eq:phip}) and (\ref{eq:phim})
respectively.
It is convenient to introduce the new variables $Q = \Phi_{+} + \Phi_{-}$ and
$\Pi = \Phi_{+} - \Phi_{-}$,
so that Eqs.~(\ref{eq:plus}) and~(\ref{eq:minus}) may be written as
\begin{eqnarray}
Q'    &=&  \Pi/\xi             \label{eq:Qdot} \\
\Pi'  &=&  Q  /\xi - 2 P_s    . \label{eq:Paidot}
\end{eqnarray}

Then, from Eqs.~(\ref{eq:statio}) we have
\begin{equation}
\frac{P'_s}{P_s} = \frac{\lambda}{2D} \frac{\Pi}{Q}
\label{eq:dlogP}
\end{equation}
which, with the aid of Eq.~(\ref{eq:Qdot}) can be integrated 
with respect to $x$, yielding
\begin{equation}
\ln P_s - r \ln Q = \mbox{const}.
\end{equation}
Thus, the steady distribution is $P_s(x) = C Q(x)^r$, where $C$ is a
normalization factor. \\

The system of ordinary differential
equations~(Eqns. ~\ref{eq:Qdot} and ~\ref{eq:Paidot}) can be regarded as an
integrable dynamical system (whose energy is conserved)
\begin{eqnarray}
Q'   &=& \Pi /\xi             \label{eq:ham1}\\
\Pi' &=&  Q  /\xi - 2 C Q^r   \label{eq:ham2}
\end{eqnarray}
describing the motion of a particle in one-dimensional potential
\begin{equation}
V(Q) = - \frac{Q^2}{2\xi} + \frac{2 C}{1+r} Q^{r+1}.
\label{eq:Pot}
\end{equation}
The spatial coordinate in the original system,
$x$,  plays the role of time in the dynamical system.
The equation of motion can be derived from the Hamiltonian,
\begin{equation}
H(\Pi,Q) = \frac{\Pi^2}{2\xi} + V(Q)
\label{eq:ham}
\end{equation}

\noindent
which is a constant of motion, $H(\Pi,Q) = E$ in the dynamical system.
Its constant value $E$ is fixed by the boundary conditions on
$P_s(x)$.  For instance, the requirement that $P_s$ be normalized for
infinite systems, implies that $Q(x=\pm\infty) = 0$ and
$\Pi(x=\pm\infty) = 0$, so that $E$ is automatically set to zero.

\begin{figure}
\centerline{\psfig{figure=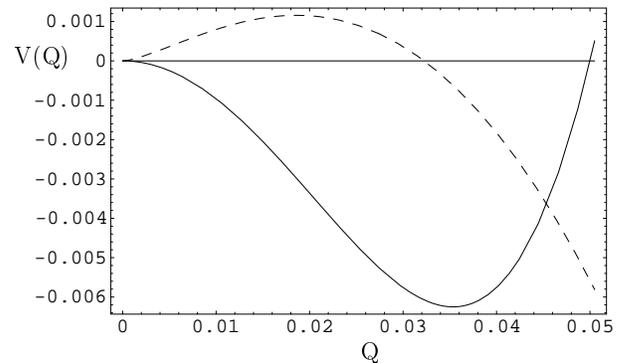,width=8cm}}
\caption{Plot of the potential $V(Q)$ for $\lambda=10,~D=1/4$ and 
$\xi=1/31$ ($r = 20/31$ -- dashed line) and $\lambda=30$~D=1/4 and 
$\xi=1/20$ ($r=3$ -- full line).}
\label{fig3}
\end{figure}

\begin{figure}
\centerline{\psfig{figure=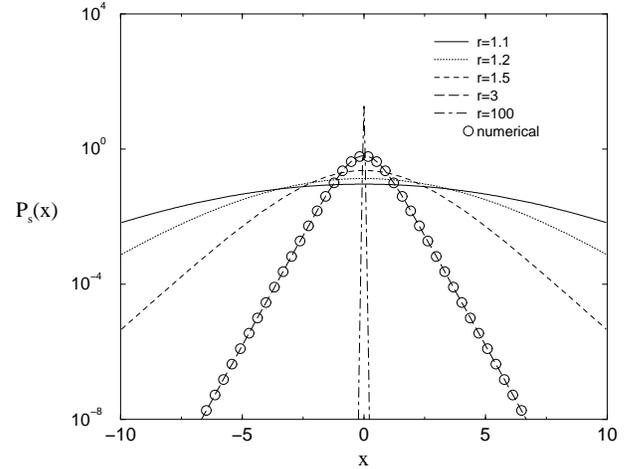,angle=270,width=8cm}}
\caption{The scaled asymptotic distribution $P_s$ for $\lambda=1$ and
several values of $r$. Solid lines refer to the exact solution
Eq.~(\ref{eq:P0new}). The open circles are obtained on numerically
integrating the differential equation (\ref{eq:Pdot}) in the large $t$
limit for $r=3$.}
\label{fig1}
\end{figure}

This mechanical analogy allows one to obtain a complete understanding
of the model with the key factor being the shape of the potential
energy (Eqn. ~\ref{eq:Pot}).  A qualitative change in the nature of
$V(Q)$ occurs (see Fig. \ref{fig3}) when $r$ crosses the value $1$,
i.e. for $ \lambda \xi = 2D$, signalling a transition in the system's
behaviour. When $r < 1$ (dashed curve of Fig. \ref{fig3}), only the
solution $Q(x)=0$ is physically acceptable and thence $P_s(x)=0$. This
corresponds to a probability distribution which spreads out and
vanishes as in standard diffusion. When $r>1$ a non-trivial trajectory
of the Hamiltonian system is possible (full line of Fig. \ref{fig3})
with $Q$ vanishing when the fictitious time $x\to\pm\infty$. This
trajectory corresponds to a non-trivial asymptotic distribution (see
Fig. \ref{fig1}). The solution for $r>1$ is obtained by integrating the
differential equation

\begin{equation} 
Q' = 1/\xi \sqrt{Q^2 - g Q^{r+1}} , 
\label{eq:dQ}
\end{equation} 

\noindent
which is obtained
by solving for $\Pi$ as a function of $Q$ (for $E=0$) and substituting
the result into Eq.~(\ref{eq:ham1}), with $g = 4 C \xi/(1+r)$. This
leads to 

\begin{equation}
P_s(x) = 
A(r)~\cosh\left[\frac{r-1}{2\xi}(x - x_0)\right]^{-2r/(r-1)},
\label{eq:P0new}
\end{equation}
where $A(r)=(r-1)\Gamma[(3r-1)/(2r-2)]/
\{2\xi\sqrt{\pi}\Gamma[r/(2r-2)]\}$ enforces 
normalization and $x_0$ is a constant which arises from the
integration of Eq.~(\ref{eq:dQ}). 

The presence of a single control parameter in the model can be deduced
by a dimensional argument. Indeed, through the rescaling: $X = x /
\xi$ and $\tau = D t/\xi^2$, the coefficient $\lambda/2$ in
Eq.~(\ref{eq:v}) becomes $r$ and the dependence on $\xi$ disappears.
Indeed when expressed in terms of $X$, the stationary state
distribution depends only on $r$.  

The change of behavior ar $r=1$ is a proper phase transition.  Indeed
it is accompanied by {\em spontaneous breaking} of the translation
symmetry $x\to x+a$ of Eq. (\ref{eq:Pdot}) for $r>1$. The location,
$x_0$, of the distribution's center is dynamically selected
and it depends, in a complex manner, on the initial 
conditions.
Furthermore, one may define a localization length, $\ell$, which
is the spread of the distribution
$P_s(x)$, and which diverges as $\ell=|1-r|^{-\nu}$
with a critical exponent, $\nu=1$, signalling the transition to the
delocalized, translationally invariant steady state. The dynamics of
relaxation to the steady state is characterized by an exponential
decay with a characteristic time diverging as one approaches the
transition. 

Another interesting feature of the system is the possibility of
spatially periodic steady-state solutions of Eq.~(\ref{eq:Pdot}) for
suitable boundary conditions, corresponding to periodic orbits of the
associated Hamiltonian systems with negative energy.  However, unlike
Eq. (\ref{eq:P0new}), such solutions are expected to be unstable to
perturbations.

\begin{figure}
\centerline{\psfig{figure=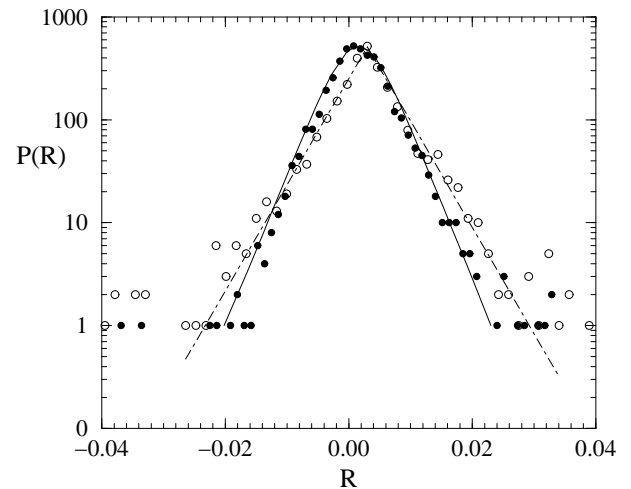,angle=270,width=8cm}}
\caption{Histograms of the monthly growth rate $R$ of two consumer price
indices
collected among $78$ Italian Provinces within the period January 1995
- June 2001.  The data are from ``Italy's National Statistical
Institute'' (ISTAT).  The filled circles refer to expenditures for
food and non-alcholic beverages, whereas the open circles show the
data for miscellaneous goods and services (see
{\tt http://www.istat.it/Anotizie/Acom/precon/indiceistat/paniere.htm} for
the entries). Both sets of data show the typical ``tent-shaped''
distribution in sharp contrast with the prediction of a Gaussian shape
of Gibrat theory.  The full and dot-dashed lines indicate a fit with
Equation~(\ref{eq:P0new}).}
\label{fig2}
\end{figure}

Many examples in which aggregation processes compete with diffusional
tendencies can also be found in economics.  Competition between
economic units may lead to effective ``peer'' pressure captured by the
drift in our model, whereas the diffusion term accounts for individual
idiosyncratic behavior.  Such a situation could arise in charitable
giving, when it is not anonymous.  A given individual would, in an
attempt to keep up with the Joneses, adjust his contribution in the
direction of higher or lower giving depending on what his peers are
doing. Of course, in this instance, the range of interactions, $\xi$,
is limited because it is not feasible for an ordinary individual to
mimic the behavior of a Bill Gates.  In other contexts, tent-shaped
distributions, which deviate sharply from a Gaussian, have been
observed for the growth rates of firms, nations' gross domestic
product (GDP) and complex organizations (\cite{Stanley} and references
therein).  We have analyzed the time series of the growth rates of
commodity prices across different Italian cities (see Fig. \ref{fig2})
and we find similar exponential tails.

The growth rate of an economic indicator $b$ -- such as the turnover of an
organization, a nation's GDP, or the price of a commodity in a city -- is
defined as
$$
R(t) = \ln \frac{b(t+1)}{b(t)}
$$
where $t$ and $t+1$ represent two consecutive time units (days,
months, years etc.).  The classical theory, due to Gibrat
\cite{Gibrat}, predicts that $R(t)$ has a Gaussian distribution.
Empirical data, however, shows a tent-shaped distribution on a
logarithmic scale. In the case of firms, Stanley {\em et al.}
\cite{Stanley,Stanley2} have proposed an explanation in terms of
the internal hierarchical
organization of non-interacting firms. 

Our model provides a different explaination in which it is the
interaction among firms which is responsible for the deviation from a
Gaussian distribution. In a way, the peer pressure mechanism can be
regarded as a {\em regression towards the mean} effect
\cite{Fridman}. This term usually refers to an explicit ``attraction''
of $R(t)$ towards a slowly varying moving average. Such an effect,
however, is unable to reproduce tent shaped distributions without
{\em ad hoc} assumptions on the precise nature of the attraction
potential. 

Here, instead, we refer to a collective phenomenon of ``regression
towards the population mean'': Each unit is attracted towards the
instantaneous population average. It is the effective interaction
among agents that generates a sort of ``feedback'' that induces the
regression towards the mean mechanism. From a microeconomic point of
view, it is reasonable to assume that the behavior of an economic
agent is influenced by other agents. For example, a firm will strive
to increase its growth rate if other peer firms grow at a higher
rate. This interaction introduces correlations in $R(t)$ that destroy
the diffusive features and produces the characteristic exponential
tails.

In summary, we have introduced and studied, both analytically and
numerically, a one-dimensional diffusion equation with nonlinear and
nonlocal features.  The asymptotic evolution of the equation leads to
the walkers being attracted toward a state which exhibits a well
defined non-Gaussian distribution characterized by exponential tails
independent of the nature of the initial condition. This mechanism of
peer pressure may provide the basis for the development of more
realistic models of self-aggregation and self-organization in
cooperative states of populations of interacting individuals\cite{ack}.


\end{multicols}  


\begin{thebibliography}{99}
\bibitem{ack} This work was supported by INFM and NASA.
\end{thebibliography}
\end{document}